\documentclass[twocolumn,preprint2]{aastex63}

\usepackage{natbib}      
\usepackage{graphicx}    
\usepackage{amsmath}
\usepackage{times}
\usepackage{amssymb}
\usepackage{stmaryrd} 
\usepackage{color}
\usepackage{bm}
\usepackage{sidecap}
\sidecaptionvpos{figure}{t}

\usepackage{comment}

\usepackage{fancyhdr}
\usepackage{epsfig}
\usepackage{float}

\usepackage{url}

\usepackage{array}
\usepackage{verbatim}
\usepackage{calc, layouts}

\hypersetup{linkcolor=red,citecolor=cyan,filecolor=cyan,urlcolor=blue}

\received{February 23, 2022}
\revised{August 1, 2022}
\accepted{August 3, 2022}

\submitjournal{ApJ}

\shorttitle{Method for Energizing MFRs Toward Eruption}
\shortauthors{Titov et al.}

\newcommand{\fb}[1]{#1}
\definecolor{Light}{gray}{.50}
\definecolor{Dark}{gray}{.20}

\definecolor{dark-red}{rgb}{0.8,0,0}
\definecolor{dark-green}{rgb}{0,0.6,0}
\definecolor{dark-blue}{rgb}{0,0,0.8}
\definecolor{dark-margenta}{rgb}{0.8,0,0.8}
\definecolor{dark-purple}{rgb}{0.45,0.2,0.65}
\definecolor{orange}{rgb}{1.0,0.6,0}
\usepackage{soul}

\newcommand{\Av}{{\bm A}}
\newcommand{\Ap}{{\bm A}_{\mathrm{p}}}
\newcommand{\Bv}{{\bm B}}
\newcommand{\Bp}{{\bm B}_{\mathrm{p}}}
\newcommand{\Bmfr}{{\bm B}_{\mathrm{MFR}}}

\newcommand{\RBSL}{{\footnotesize R}BS{\footnotesize L}}

\begin{document}
%

\title{A Magnetogram-matching Method for Energizing Magnetic Flux Ropes Toward  Eruption} 

\correspondingauthor{Viacheslav S. Titov}
\email{titovv@predsci.com}

\author[0000-0001-7053-4081]{V. S. Titov}
\affiliation{Predictive Science Inc., 9990 Mesa Rim Road, Suite 170, San Diego, CA 92121, USA} 

\author[0000-0003-1759-4354]{C. Downs}
\affiliation{Predictive Science Inc., 9990 Mesa Rim Road, Suite 170, San Diego, CA 92121, USA} 

\author[0000-0003-3843-3242]{T. T\"{o}r\"{o}k}
\affiliation{Predictive Science Inc., 9990 Mesa Rim Road, Suite 170, San Diego, CA 92121, USA} 

\author[0000-0003-1662-3328]{ J. A. Linker}
\affiliation{Predictive Science Inc., 9990 Mesa Rim Road, Suite 170, San Diego, CA 92121, USA} 

\begin{abstract}
We propose a new ``helicity-pumping'' method for energizing coronal equilibria that contain a magnetic flux rope (MFR) toward an eruption.
We achieve this in a sequence of magnetohydrodynamics relaxations of small line-tied pulses of magnetic helicity, each of which is simulated by a suitable rescaling of the current-carrying part of the field.
The whole procedure is ``magnetogram-matching'' because it involves no changes to the normal component of the field at the photospheric boundary.
The method is illustrated by applying it to an observed force-free configuration whose MFR is modeled with our regularized Biot--Savart law method.
We find that, in spite of the bipolar character of the external field, the MFR eruption is sustained by two reconnection processes.
The first, which we refer to as breakthrough reconnection, is analogous to breakout reconnection in quadrupolar configurations.
It occurs at a quasi-separator inside a current layer that wraps around the erupting MFR and is caused by the photospheric line-tying effect.
The second process is the classical \fb{flare} reconnection, which develops at the second quasi-separator inside a vertical current layer that is formed below the erupting MFR.
Both reconnection processes work in tandem with the magnetic forces of the unstable MFR to propel it through the overlying ambient field, and their interplay may also be relevant for the thermal processes occurring in the plasma of solar flares.
The considered example suggests that our method will be beneficial for both the modeling of observed eruptive events and theoretical studies of eruptions in idealized magnetic configurations.
\end{abstract}


\section{Introduction
 \label{s:intro}}

In this paper we focus on one of the key questions  of modeling observed or idealized solar eruptions: Given a pre-eruptive magnetic configuration, what is an efficient way to bring it to a loss of equilibrium or to make it unstable under imposed observational constraints?

One of the most popular approaches relies on the magnetic flux cancellation process.
\citet{vanBall1989} suggested that this process can transform an idealized sheared magnetic arcade (SMA)  into an unstable configuration.
The flux cancellation is driven by photospheric flows converging to the polarity inversion line (PIL) of the SMA, where the magnetic diffusion is locally enhanced.
This process gradually forms a magnetic flux rope (MFR) above the PIL; the MFR can eventually become unstable if the flux cancellation lasts long enough. 

This idea has been successfully extended and applied to more realistic magnetic configurations \citep[e.g.,][]{Amari2000, Linker2001,linkeretal2003,  Zuccarello2012, Mikic2013a, Torok2018, Hassanin2022}.
Because the normal component of the photospheric field is canceled during this process, the initial distribution of this component may be modified such that the observed distribution (obtained from magnetograms) is roughly matched after the cancellation process.
However, in order for this to succeed, several iterations may be required because the critical flux to be canceled for triggering the eruption is unknown a priori.

The situation appears to be even more complicated if one takes into account that in reality the critical flux may often be localized around a small segment of the PIL whose determination requires a rather sophisticated method \citep{Kusano2020}.

The inherent challenges of using flux cancellation as a tool for triggering eruptions have motivated us to develop a simpler approach that is still physics-based.
From a general point of view, it is clear that, in order to cause an eruption in a given configuration, one needs somehow to raise its magnetic energy to a critical level where a stable equilibrium can no longer exist.
\fb{In reality, such a critical state is reached in a quasi-static evolution of the coronal magnetic field, which is driven by the slow dynamics of the dense photospheric plasma.
A full modeling of such an evolution is beyond the scope of this study.
Instead, we assume that a slightly subcritical magnetic equilibrium is already present, and that this equilibrium requires only some additional energization to reach a critical state.}
For an accurate modeling of observed events, the following constraints on the energization procedure are desirable: 
1) The magnetic structure of the configuration must remain similar to the initial one during the energization process;
2) The final distribution of the normal field at the boundary should precisely match the distribution \fb{and} corresponding observations 
\fb{of the source-region configuration}
at the onset of the eruption.

By using that distribution as a boundary condition for the initial equilibrium and preserving it during the energization process, the first constraint is likely met if the critical value of the magnetic energy is relatively close to the initial one.
This condition is best met if one uses as many observational constraints as possible when modeling the initial equilibrium.

In particular, a rigorous constraint on the field-line connectivity in the modeled configuration serves well for this purpose, as was demonstrated by the field models constructed with the MFR insertion method \citep[e.g.,][]{vanBall2004,Savcheva2009a}.
Models of this type become even more accurate if the shape and total current of the MFR are suitably optimized \citep{Titov2021}.
Such an optimized configuration is used here as the initial equilibrium in an example application.
As we shall see, its free magnetic energy is smaller 
than the critical one by only one-third, which appears to be sufficient for the similarity constraint mentioned above to be fulfilled.

Based on these considerations, we propose here a new method for energizing pre-eruptive configurations toward eruptions.
Section \ref{s:method} describes our method in detail.
Section \ref{s:example} presents its application to a bipolar pre-eruptive configuration.
Section \ref{s:sum} summarizes the obtained results.

\section{Helicity-Pumping Method
 \label{s:method}}

We energize a given pre-eruptive configuration, which typically consists of an MFR embedded in a background potential field in a sequence of cycles, each of which employs two operations: first, a suitable rescaling of the current-carrying component of the total magnetic field and, second, the relaxation of this field toward equilibrium via line-tied zero-$\beta$  magnetohydrodynamics (MHD) simulations, where thermal pressure and gravity are neglected.

More precisely, at each cycle of the sequence, the total equilibrium field $\Bv$ is decomposed as $\Bv = \Bp + \Bmfr$, where 
$\Bp$ is the potential field derived from a given normal or radial component of the magnetic field at the photospheric boundary, while
$\Bmfr \equiv \Bv-\Bp$ is the remaining nonpotential or MFR field with a vanishing normal component at the boundary.
We then multiply $\Bmfr$ by a factor $1 + \varepsilon$ (with a small positive increment $\varepsilon$) to obtain a new total 
\fb{(``perturbed'')} field
\begin{eqnarray}
\tilde{\Bv}
=
\Bp
+
(1 + \varepsilon)
\,
\Bmfr
\,.
	\label{Bnew}
\end{eqnarray}
This operation leaves the normal component at the boundary unchanged, while raising the electric current \fb{($\sim (1+\varepsilon)$) and free} magnetic energy \fb{($\sim (1+\varepsilon)^2$)} of the configuration \fb{(see Section \ref{ss:energy}).} 
\fb{Also}, as a result \fb{of this operation}, the magnetic forces become unbalanced in the volume to an extent that depends on the chosen value of $\varepsilon$.
Therefore, we relax these forces in a zero-$\beta$  MHD simulation  under line-tying boundary conditions, which preserves the normal component of $\tilde{\Bv}$ at the boundary.
This relaxation completes one cycle of the procedure and provides us with a new, slightly more energized $\Bv$ and a somewhat different tangential component at the boundary.
By repeating this cycle, we eventually raise the free energy of the configuration to the level where a stable equilibrium can no longer exist and an eruption occurs.
\fb{We expect a stepwise growth of the helicity (``helicity pumping'') in this procedure, bearing in mind the nearly monotonic dependence between the free magnetic energy and relative magnetic helicity in active regions (see \citet{Tziotziou2014} and references therein).
From the perspective of boundary-driven evolution methods, the increase of magnetic energy and helicity in our resulting sequence of equilibria occurs due to changing the tangential component of the magnetic field at the boundary, while the corresponding normal component remains unchanged by construction.
}

The occurrence of an eruption after a sufficient number of cycles is \fb{to be expected.
Indeed, the perturbed field defined by Eq. \ref{Bnew} can alternatively be represented as}
\begin{eqnarray}
\tilde{\Bv} \equiv  (1 + \varepsilon)\, \Bv - \varepsilon \Bp\, ,
	\label{Bnew2}
\end{eqnarray}
\fb{which} is simply
\fb{a previously relaxed}
field
\fb{amplified by the factor $(1+\varepsilon)$}
\fb{minus the fraction $\varepsilon$ of the potential field component.
This component acts inside the MFR as a strapping field;}
\fb{therefore, its reduction implies that the MFR will slightly rise and expand over each cycle until, in all likelihood, it reaches 
a height where it}
becomes unstable (Section \ref{ss:energy}).
Whether then the helical kink or the torus instability is realized depends on the specific properties of the MFR and the \fb{ambient}
potential field 
\citep[e.g.,][]{torok04,kliem06}\fb{, which remains a subject of intense research
\citep[e.g.,][]{Hassanin2016, Jing2018}}.

We note that \fb{the energization of pre-eruptive  configurations by} our helicity-pumping method does not pose a requirement on 
how the electric current
\fb{should be}
distributed in the initial equilibrium.
\fb{The method} can be applied to magnetic configurations with a current structure more complex than that of a single MFR, as demonstrated below.
\fb{However, in order to reach an unstable equilibrium with
this method, the initial current structure 
likely has to be
sufficiently elongated.}
\begin{figure*}[ht!]
\plotone{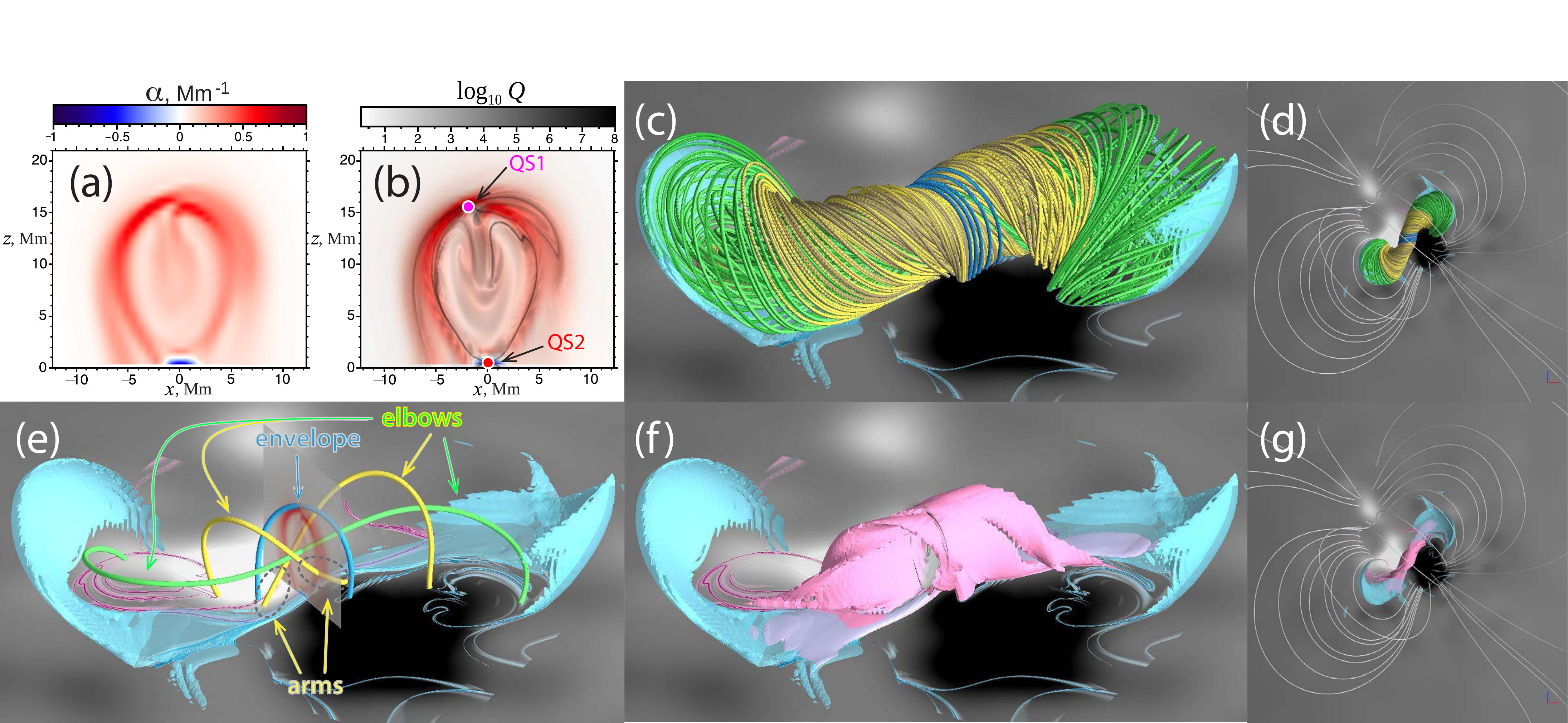}
\caption{
The initial pre-eruptive equilibrium: maps of $\alpha$ (panel (a)) and $\log_{10}\,Q$ (panel (b)) in the central cross section of the configuration whose magnetic field lines and current layers are shown in panels (c) and (f) (side view) and panels (d) and (g) (top view), respectively. Field lines of the MFR and SMA are colored in green and yellow, respectively.
Isosurfaces $j/j_{\mathrm max} = 0.438$ (magenta) and $\alpha/\alpha_{\mathrm min}=0.079$ (semitransparent cyan) show the corresponding layers of direct and return currents.
Panel (e) depicts representative field lines corresponding to observed morphological features \citep[see][]{Titov2021}.
The photospheric $B_r$ distribution is shown by gray shading from white ($B_r>0$) to black ($B_r<0$); the overlaid high-$Q$ lines colored in magenta (if $B_r>0$) and cyan (if $B_r<0$) outline the footprints of the MFR and SMA. 
The two quasi-separators (QS1 and QS2) described in this section are indicated in panel (b).
\label{fig1}}
\end{figure*}

\fb{It is worth noting that, by choosing a negative value for $\varepsilon$, one can model a ``reverse pumping'' or ``deflation'' of the helicity (and free energy) with our method. This could prove useful in situations where, for example, a constructed configuration matches the observed magnetogram, but the initial MFR is somewhat over-energized and therefore immediately unstable. By applying reverse pumping to such a configuration, a critical
state can be reached in a ``backward evolution,'' and the eruption onset can then be modeled more accurately.}

\section{Illustrative Example
 \label{s:example}}

\subsection{Initial Equilibrium
 \label{ss:init}}

To see how our helicity-pumping method works in practice, we apply it to the pre-eruptive configuration of the 2009 February 13 CME event \citep[see, e.g.,][]{patsourakos09,downs21} that we recently constructed using our upgraded \RBSL\ method in \citet[]{Titov2021}, which is based on the so-called regularized Biot--Savart laws \citep{Titov2018}.
Figure \ref{fig1} summarizes the basic properties of this sigmoidal pre-eruptive equilibrium.
In particular, panel (a) shows the distribution of the force-free parameter $\alpha = \Bv \cdot \nabla \times \Bv / \Bv^2$ in the central cross section of the modeled configuration.
Panel (b) compares this distribution with the corresponding map of the squashing factor $Q$ \citep{Titov2002, Titov2007a} whose high-$Q$ curves are cross-sectional intersections of so-called quasi-separatrix layers \citep[QSLs;][]{Priest1995, Demoulin1996}.
One can see that the electric current is concentrated in layers that closely follow these high-$Q$ curves.
The corresponding QSLs and current layers in the volume (see panels (c)--(g)) serve as boundaries for the magnetic building blocks of the configuration, which include an MFR (green field lines) nested into an SMA (yellow field lines).

As described in \cite{Titov2021}, this force-free equilibrium was obtained via a zero-$\beta$ MHD relaxation of an initial configuration that contained an \RBSL\ MFR whose shape and total current were previously optimized to minimize unbalanced magnetic forces in the volume.
The axis path of the optimized MFR was mirrored about the photospheric boundary to provide a closure of the current and to match the observed magnetogram.
The line-tying conditions applied at the photospheric boundary preserve the magnetogram during the relaxation.

Panels (f) and (g) present the electric-current structure of the relaxed configuration.
An isosurface of the modulus of the current density is colored in magenta, while the cyan surface depicts an isosurface of a negative $\alpha$ value.
For our further considerations, it is important to note that the magenta current layer wraps around the MFR (green field lines in panel (c)).
It is also noteworthy that the overall field-line structure agrees well with
morphological features that have been observed prior to, or early on, in the eruptions of sigmoids \citep[e.g.,][]{Moore2001}.
Panel (e) shows representative field lines that can be associated with 
the observed ``elbows,'' ``arms,'' and ``envelop'' features described by those authors.

The $Q$-map in panel (b) has two X-type intersections of the high-$Q$ curves in the considered cross section. Those intersections define local maxima of $Q$, which are designated by small magenta and red circles.
The field lines passing through these maxima are so-called quasi-separators (QSs), which are a geometrical generalization of topological features such as separators and X-lines \citep{Titov2002, Titov2007a}.
\fb{The appearance of QS1 (magenta) is, in contrast to QS2 (red), 
not expected, and worthwhile of a thorough study in the future.
Here we provide only a few preliminary considerations.} 

\fb{We think that QS1 appears in our configuration due to mirror images of the axial and azimuthal currents of the MFR about the photospheric boundary.}
\fb{The axial} image current serves to compensate the normal field of the MFR at the boundary \fb{outside of the MFR footprints} \citep[][]{Isenberg2007}.
\fb{ By inverting the sign in the subphotospheric part of the azimuthal \RBSL\ vector potential \citep[see Eq. (9) in][]{Titov2021}, one additionally obtains this compensation inside the MFR footprints.}

\fb{The MFR axis path and its subphotospheric image} follow the PIL of the bipolar magnetogram.
The \fb{axial} image current, therefore, creates at the boundary, in the vicinity of the PIL, a bipolar distribution of the normal-field component that is opposite to the original distribution.
Thus, the sum of the field induced by the \fb{axial} image current and the potential field $\Bp$  has a quadrupolar structure whose QS serves as an approximate axis path for the MFR.
In the resulting total field $\Bv$, which takes into account also the field of the MFR current, this QS 
\fb{splits} into QS1 and QS2, which are\fb{, respectively,} located above and below the MFR  \fb{in the central cross section shown in panel (b).
QS1 is adjacent to each of the MFR, SMA, and envelope-field regions and accommodates the partition of magnetic flux between them.
QS1 is located at the place of contact of the current layers that border the MFR and SMA and follow the corresponding QSLs.
It is natural then to assume that QS1 and the SMA can evolutionarily be formed 
as a result of magnetic reconnection between the axial MFR field and the overlying envelope field.}

Note that QS2 is present in similar, bipolar MFR configurations even in the absence of an image current, provided that the MFR apex is located high enough above the boundary.
Its presence is a result of the superposition of the bipolar and MFR-current fields \citep{Demoulin1996a}.
For our configuration, the introduction of the \fb{axial} image current into the configuration shifts the location of this QS2 \fb{toward} the boundary.

%
\begin{figure*}[ht!]
\plotone{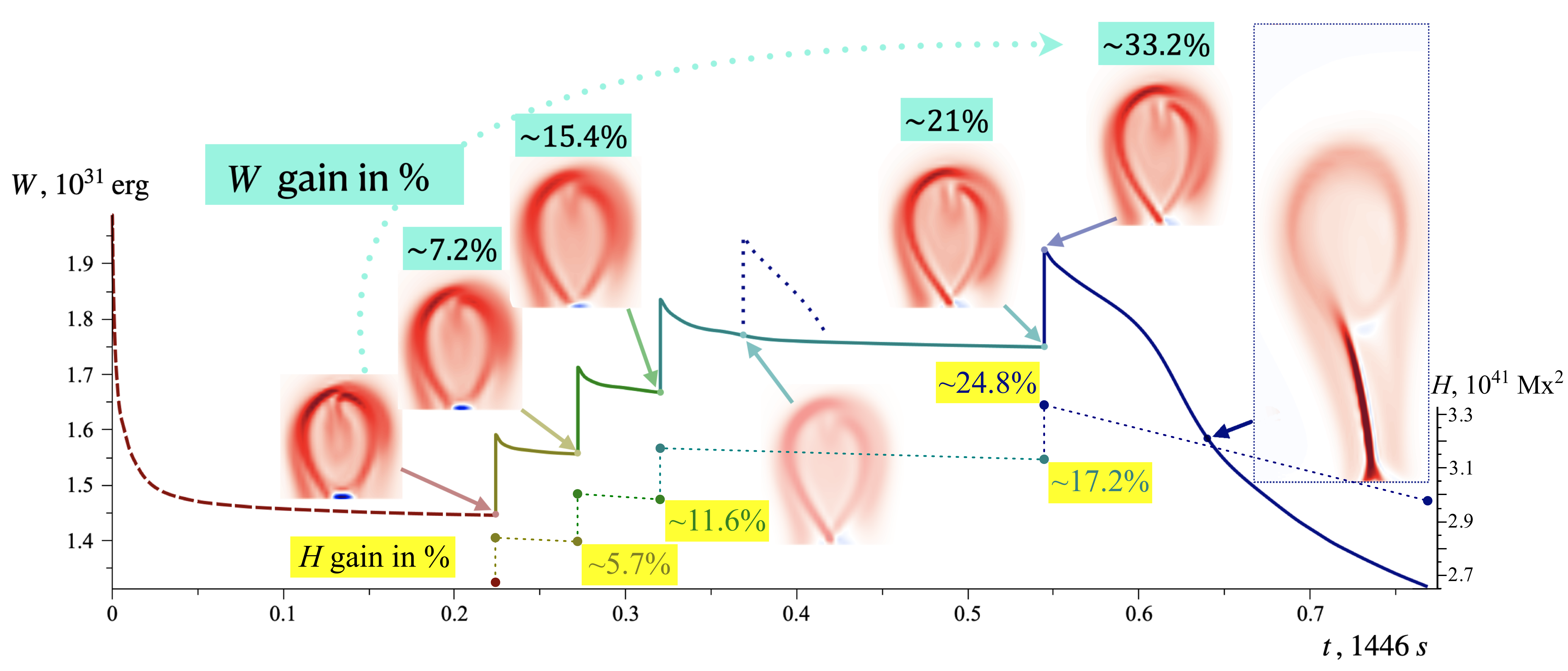}
\caption{
The evolution of the free magnetic energy $W$ \fb{(its scale is shown at the axis on the left)} of our pre-eruptive configuration (see Figure \ref{fig1}) during the energization by our helicity-pumping method  (solid lines).
The distribution of $\alpha$ in the central cross section of the configuration is shown at the beginning or end of each cycle to show how the current structure evolves during this process.
For comparison, the dashed brown curve presents the decay of the free energy during the initial relaxation of the optimized \RBSL\ MFR configuration toward the pre-eruptive configuration.
The relaxation phase of the third cycle is prolonged from $t\approx 0.37$ to $t\approx 0.55$ to prove that the preliminary last cycle (dotted line) is indeed destabilizing.
\fb{Small circles connected by dotted lines show a stepwise increase of the relative helicity $H$ (its scale is shown at the axis on the right) in our energization procedure.}
\label{fig2}}
\end{figure*}
\subsection{Energization Process
 \label{ss:energy}}

Let us demonstrate now how our helicity-pumping method works by applying it to the above pre-eruptive equilibrium.
\fb{
The relative helicity $H$ of two fields $\Bv$ and $\Bp$, proposed by \citet{Berger1984}, is useful to transform to \citep{Finn1985}
\begin{eqnarray}
H = \int_V {\mathrm d}V\, (\Av + \Ap)\cdot(\Bv-\Bp)
\,,
	\label{H}
\end{eqnarray}
which, because we advance the vector potential directly in the MHD calculation, can be calculated at different times during
our energization procedure.
By construction, the radial 
component 
of $\Bv$ and of the
potential field, $\Bp=\nabla\times\Ap$, $r=R_{\sun}$ were kept equal
at the lower boundary, $r=R_{\sun}$, during our simulation.
Along with the imposed vanishing boundary conditions for these components at the upper boundary $r=10\,R_{\sun}$, this ensures that the resulting values of $H$ are  independent of the gauge of the vector potentials $\Av$ and $\Ap$.
}

For all cycles of the energization, we use the same incremental fraction $\varepsilon=0.05$, which leads, after three cycles, to an increase of \fb{the relative helicity $H$ by $\sim 17.2\%$ and of} the free magnetic energy $W$ by $\sim 21\%$ (see Figure \ref{fig2}).
The rescaling \fb{of $\Bmfr$} itself\fb{, given by Eq. (\ref{Bnew}),} yields \fb{$\sim 6\%$ and} $\sim 10\%$ per cycle for the increment of \fb{$H$ and} $W$, \fb{respectively.
The latter grows approximately $\propto (1+\varepsilon)^2$, as expected from}
the quadratic dependence of $W$ on the total electric current.

The self inductance of the configuration is not affected by the rescaling, but it is changed in the subsequent MHD relaxation.
What is likely more important for the evolution of $W$, however, is that a fraction of $W$ is converted during the relaxation into  kinetic energy, which largely dissipates toward the end of each cycle.
In addition, substantial ohmic dissipation of the free energy has to occur in the current layers of the configuration.
Altogether, these effects of the relaxation result in a drop of $W$ by $\sim 3\%$ \fb{and $H$ by $\lesssim 1\%$} per cycle.
\fb{As expected, $H$ decays at each cycle during the relaxation much slower than $W$.
A significant decay of $H$ by $\sim 11\%$ 
occurs only during the fourth cycle when a strong reconnection develops  {in current layers that rapidly grow in size} in the configuration.
Given the explicit resistivity term and finite, nonuniform mesh spacing in the model, we suspect that the effective magnetic Reynolds number is not high enough to perfectly conserve helicity during this phase of the eruption.
}

\fb{During this fourth cycle},
$W$ does not \fb{relax toward a constant value} (dotted line in Figure \ref{fig2}), which indicates a destabilization of the configuration.
To be sure that this destabilization is not the result of a residual imbalance of magnetic forces, we extended the relaxation of the configuration obtained after the third cycle for a longer time and then resumed the fourth cycle.
The resulting gain in $W$ reached $33\%$ in the fourth cycle after rescaling, and an MFR eruption occurred.
The initial decay of $W$ in this resumed fourth cycle is similar to the one we obtained in the preliminary version of this cycle, when we started with a slightly nonrelaxed equilibrium.
If desired, a more precise determination of the transition to an unstable configuration, and the corresponding estimate for the critical value of $W$, can be obtained iteratively by repeating the last two cycles with a suitably adjusted $\varepsilon$.

The maps of $\alpha$ in the central cross section of the configuration at the beginning or end of each cycle give us an idea of how the structure of the configuration evolves during the energization.
The MFR and SMA slightly expand and stretch in the vertical direction from cycle to cycle but otherwise remain similar to the initial state that is illustrated in detail in Figure \ref{fig1}.
By comparing to the $\alpha$- and $Q$-maps at later times, we can see that the current remains concentrated along QSLs throughout all cycles of the process.
The latter is true for the eruption phase as well, which will become clear in Section \ref{ss:rec}.

The dashed brown curve in Figure \ref{fig2} shows the evolution of $W$ during the preliminary relaxation of the optimized \RBSL\ MFR configuration toward our starting equilibrium shown in Figure \ref{fig1}.
It is interesting that the free energy of this configuration (at $t=0$) was even a bit higher than the initial value of $W$ in the (eruptive) fourth cycle.
Nevertheless, the optimized MFR did not erupt but rather transformed into our starting equilibrium \citep[see Section \ref{ss:init} and, for more details,][]{{Titov2021}}.
Such a difference in the behavior of two MFR configurations is probably due to distinct current (and, hence, Lorentz force) distributions: in the optimized MFR, the current was distributed over its cross section rather than concentrated in layers.

\begin{figure*}[ht!]
\plotone{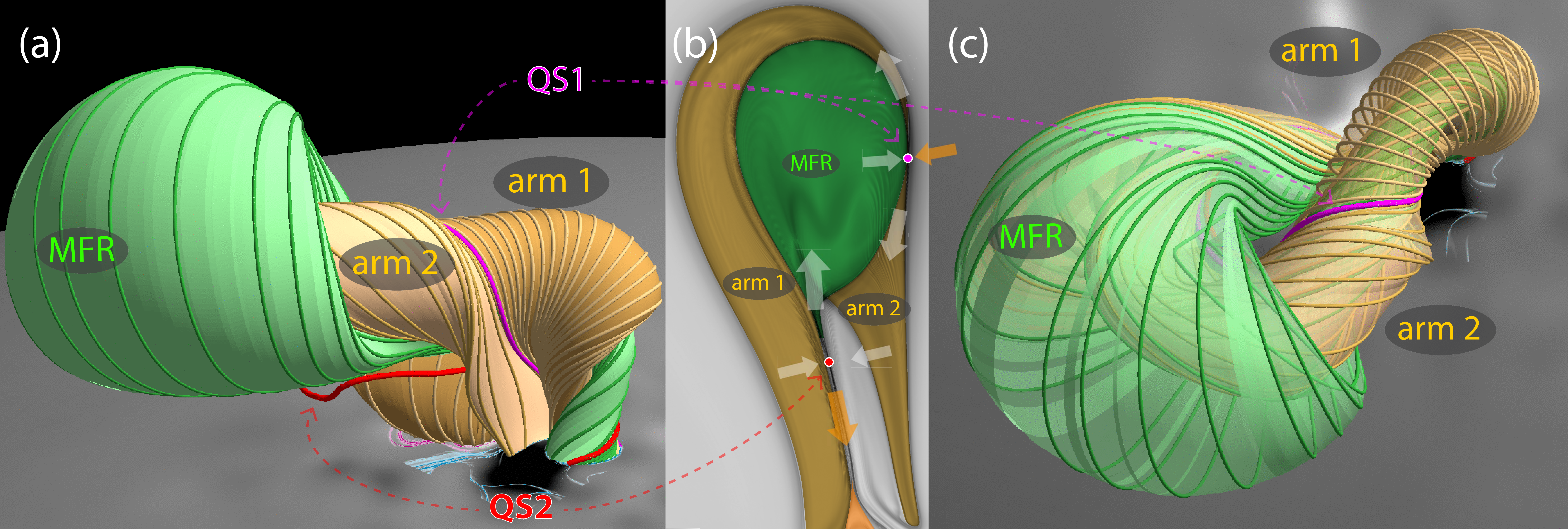}
\caption{
Side (a) and top (c) views of the magnetic field structure during the $W$ evolution presented in Figure \ref{fig2}, shown here at $t=0.64$. The
shown magnetic surfaces that bound the MFR and its ``arms'' and the quasi-separators QS1 and QS2 are derived from the $Q$-map (b) in the central cross section of the configuration (cf. Figure \ref{fig1}(e)).
\label{fig3}}
\end{figure*}

\vspace{1.0cm} 
\subsection{Reconnection at the Initial Stage of Eruption
 \label{ss:rec}}

The nontrivial structure of the initial equilibrium, with two QSs located inside the current layers (Section \ref{ss:init}), remains very similar during the energization process.
During the eruption, these topological features change their locations 
within the expanding and stretching MFR/SMA structure, but remain always inside the evolving current layers.
Their presence has a profound impact on the eruption process, due to magnetic reconnection occurring in their vicinity.

Figure \ref{fig3} shows the locations of the QSs within the erupting structure during the initial stage of the eruption and illustrates their impact.
Panel (b) presents a $Q$-map in the same cross section as in Figure \ref{fig1}(e) at the time when the MFR has reached three times its initial height. Panels (a) and (c) show side and top views, respectively, of the corresponding field-line structure derived from this $Q$-map.
The high-$Q$ curves (dark gray) separate or outline the building blocks of the configuration: the MFR (shaded in green), the vertical current layer beneath the rope, and two adjacent flux tubes (shaded in yellow), which we call arms, following \citet[][]{Moore2001}.
Initially, these arms belong to the SMA that encloses the MFR (yellow field lines in Figure \ref{fig1}(c)).

During the course of the eruption, the magnetic flux within the arms is replenished by reconnection at QS1 \fb{(see Figure \ref{fig3})}, which is located in the current layer that wraps around the MFR.
Simultaneously, the reconnection reduces the fluxes of the MFR and of the overlying envelope field.
We refer to this reconnection as {\it breakthrough} reconnection, which is analogous to breakout reconnection \citep{Syrovatskii1982, Antiochos1999}, except that it develops in a bipolar rather than quadrupolar configuration.
As explained in Section \ref{ss:init}, the  magnetic field external to the MFR has actually a ``stealth'' quadrupolar character in our bipolar configuration because of the fictitious image current below the boundary.
And, due to the photospheric line-tying conditions, the initial bipolar distribution of the normal field remains unchanged during the eruption.
This means that the image current evolves such that it continuously compensates the varying normal-field component that would otherwise be produced by the erupting coronal current at the boundary.
In other words, the line-tying conditions support the quadrupolar character of the field external to this current, and hence the presence of QS1 in the configuration during the eruption.

Note that, as explained by \citet{Isenberg2007}, the image current is not intended to resemble the actual currents below the photospheric boundary.
Physically, it is constructed to represent the surface currents that are induced in response to the line-tied MHD evolution of the coronal current.
For the above considerations, it is only important that the surface currents and the image current generate \fb{identical contributions to the potential field} in the corona.

QS2 is located in the vertical current layer beneath the MFR, and it is responsible for the classical 
\fb{flare} reconnection that occurs across this layer.
This reconnection merges the lower sections of the arms and reduces their flux, while raising the flux of the MFR and the developing flare arcade (shaded in orange in Figure \ref{fig3}(b)).

Both reconnection processes seem to operate simultaneously 
\fb{right from}
the onset of our modeled eruption.
The arrows in Figure \ref{fig3}(b) indicate inflows and outflows of the magnetic flux at the reconnection sites, where white arrows designate the aforementioned recirculation of the magnetic flux between the erupting MFR and the arms.
Although such a recirculation has yet to be properly quantified, it is already clear that it 1) prevents a disintegration of the MFR body during the initial stage of eruption, and 2) makes two reconnection processes work in tandem, helping to propel the erupting MFR through the overlying envelope field.
The recirculation is not a closed process, as it also involves the inflow of the envelope-field flux at QS1 and the outflow of the reconnected flux at QS2 downward to the flare arcade.
This inflow and outflow are indicated in Figure \ref{fig3}(b) by orange arrows.

The potential background field $\Bp$ at the position of the left leg of the initial MFR is a bit weaker, which results in a slightly larger expansion of this leg (Figure \ref{fig1}).
As the eruption proceeds, the asymmetry is amplified, which manifest in a strong sideward bulging and expansion of this leg (Figure \ref{fig3}(a) and (c)).

By definition, a QS belongs to an X-type intersection of two QSLs that form a so-called hyperbolic flux tube \citep[HFT;][]{Titov2002,Titov2007a}.
Reconnection typically  occurs in a pinched subvolume of the HFT that has an enhanced current density \citep[e.g.,][]{Savcheva2015}.
Thus, by tracking down the QSs in our current layers, we actually identify pinched HFTs, where magnetic reconnection takes place.
With this in mind, note that two types of reconnections similar to ours have recently been identified  in a simulation by \citet{LiuT2021} in two HFTs of an erupting, idealized quiet-sun prominence configuration. 
The eruption resulted from the emergence of a symmetric toroidal MFR at the lower boundary of a helmet-streamer configuration \citep{Fan2017}.
The HFTs were formed subsequently, right after the onset of the eruption,
first above and then below the MFR \fb{in the corresponding current layers}.
This is consistent with our 
above explanation that the line-tying effect has to lead to the formation of two QSs (and the corresponding HFTs) near the MFR during an eruption.

\

\section{Summary and Discussion
 \label{s:sum}}

We have proposed a new method for gradually energizing magnetic equilibria toward eruption while preserving the normal magnetic field at the photospheric boundary.
This ``helicity-pumping'' technique overcomes some inherent challenges present in other approaches for modeling eruptions. For example, with a boundary driving approach that modifies the surface flux distribution using flows, it can be difficult to tune the parameters a priori such that the observed flux distribution is matched at the time of eruption onset. Similarly, out-of-equilibrium MFR insertion methods \citep[e.g.,][]{manchester08,lugaz11,Jin2017} may substantially modify the surface field, thus overestimating the free energy and/or producing erupting structures that could never have been supported by the \fb{source active region} 

In our method, the configuration is energized in a series of cycles, each of which consists of a small pulse of magnetic helicity and a subsequent short MHD relaxation, both performed under line-tying boundary conditions.
The helicity pulse is realized by a suitable rescaling of the nonpotential part of the previous cycle's field with a vanishing normal component at the boundary.
At first sight, such pulsation might appear unphysical, because it is created in the whole volume instantaneously rather than gradually via evolving the field at the boundary.
However, we should remember that in reality there are always fluctuations or
perturbations of the magnetic field in the low corona, whose time-averaged impact on the magnetic energization process might not be negligible.
In this respect, the proposed method can be seen 
as an approximate, but practical way to incorporate this effect.
Indeed, each helicity pulse in our example application induces \fb{an out-of-equilibrium}
\fb{perturbation of the configuration}  whose subsequent line-tied MHD relaxation \fb{is accompanied by several decaying oscillations of the structure and waves propagating 
within and out of it.
The relaxation eventually}
leads to an accumulation of the current density and related magnetic stress along the QSLs of the configuration.
This behavior is well consistent with the QSL concept \citep[][]{Priest1995, Demoulin1996, Demoulin1996a, Titov2007a}.
However, to the best of our knowledge, there are no other methods that emulate this effect in such a simple and efficient manner\fb{, which therefore makes our method unique}. 

The application of the helicity-pumping method to an observed sigmoidal pre-eruptive equilibrium demonstrates the uniqueness, efficiency, and importance of our approach.
One of the most interesting results obtained in this application relates to the eruption process in a bipolar configuration.
We have found that, in addition to the standard \fb{flare} reconnection, the initial evolution of the eruption is guided by ``breakthrough reconnection,'' which is analogous to breakout reconnection in quadrupolar configurations.
Due to these two reconnections, the magnetic flux recirculates between the MFR and the ``arms'' embracing it, which helps to preserve the integrity of the MFR.
The arms are sustained by the breakthrough reconnection of the MFR and the overlying envelope magnetic field.

The concurrent operation of both reconnections during an eruption implies important consequences for the embedded coronal plasma.
First, a part of the magnetic energy released by the breakthrough reconnection must locally heat the plasma.
The thermal fluxes flowing from this region downward to the boundary, in turn, should evaporate plasma into the new flux tubes added to the arms.
Therefore, when the arms then enter the vertical current layer during the \fb{flare} reconnection, they will likely contain a denser plasma than the surrounding quiet corona.
This process may be important for understanding the ``hot'' or post-impulsive phase of solar flares, where the reconnection process likely has a quasi-steady character \citep[e.g.,][]{Forbes2018}.

Indeed, self-consistent \fb{two-dimensional} estimates of the characteristics of reconnecting current layers under these quasi-steady conditions \citep{Somov1985a, Somov1985b} convincingly show that the density of plasma inside such layers can be only a few times larger than outside in the inflow region.
This result is very robust because it was derived from a full set of conservation laws written in the integral form and therefore
does not depend on different assumptions used for the anomalous resistivity in the current layers.
\fb{Such a modest compression of the flare plasma in the current layers is due to the fact that their cooling at temperatures around $\sim 10^7\,{\mathrm K}$ is primarily carried out by anomalous thermal fluxes and thermal enthalpy outflows, while the radiative loses in the current layers are negligibly small at these conditions.}
However, \fb{to sustain the rate of energy release required for the post-impulsive phase of flares,} the plasma density \fb{in the layers should reach values around $\sim 10^{10}\,\mathrm{cm}^{-3}$, which is} one or two orders of magnitude larger than in the \fb{surrounding coronal plasma} \citep[see, e.g.,][]{Priest1982}. 
Thus, if the vertical current layer below an erupting MFR is a source of hot flare plasma, then there must be an additional physical process that raises the inflow density of the plasma to the \fb{required} levels.
It appears that \fb{evaporation of chromospheric plasma by} our breakthrough reconnection is a natural candidate for this process.

\fb{Alternatively, the problem of the required high plasma density in flare current layers may be solved 
by invoking an essentially three-dimensional character of plasma flows in the layers, which is not included in the above-mentioned two-dimensional estimates.
In addition to the transverse stagnation-type flows in the layers, strong counter-directed flows may be induced along the layers by their own plasma evaporation processes, which could supply the layers with the required high-density  plasma.}

This example application demonstrates the usefulness of our helicity-pumping method for both the modeling of realistic CME events and theoretical studies of eruptions in idealized configurations. 


\acknowledgments

\fb{We are very thankful to Pascal D\'emoulin, Terry Forbes, Bernhard Kliem, and the anonymous referee for their valuable 
comments that helped us to improve this paper.}
This research was supported by NASA grants 80NSSC20K1317, 80NSSC20K1274, and 80NSSC19K0858; by NASA contract 80NSSC19C0193; by NASA/NSF grant 80NSSC20K0604; and by NSF grants AGS-1923377 and ICER1854790. Computational resources were provided by NSF's XSEDE and NASA's NAS.



\begin{thebibliography}{}
\expandafter\ifx\csname natexlab\endcsname\relax\def\natexlab#1{#1}\fi
\providecommand{\url}[1]{\href{#1}{#1}}
\providecommand{\dodoi}[1]{doi:~\href{http://doi.org/#1}{\nolinkurl{#1}}}
\providecommand{\doeprint}[1]{\href{http://ascl.net/#1}{\nolinkurl{http://ascl.net/#1}}}
\providecommand{\doarXiv}[1]{\href{https://arxiv.org/abs/#1}{\nolinkurl{https://arxiv.org/abs/#1}}}

\bibitem[{{Amari} {et~al.}(2000){Amari}, {Luciani}, {Mikic}, \&
  {Linker}}]{Amari2000}
{Amari}, T., {Luciani}, J.~F., {Mikic}, Z., \& {Linker}, J. 2000, \apjl, 529,
  L49, \dodoi{10.1086/312444}

\bibitem[{{Antiochos} {et~al.}(1999){Antiochos}, {DeVore}, \&
  {Klimchuk}}]{Antiochos1999}
{Antiochos}, S.~K., {DeVore}, C.~R., \& {Klimchuk}, J.~A. 1999, \apj, 510, 485,
  \dodoi{10.1086/306563}

\bibitem[{{Berger} \& {Field}(1984)}]{Berger1984}
{Berger}, M.~A., \& {Field}, G.~B. 1984, Journal of Fluid Mechanics, 147, 133,
  \dodoi{10.1017/S0022112084002019}

\bibitem[{{D{\'e}moulin} {et~al.}(1996{\natexlab{a}}){D{\'e}moulin}, {Henoux},
  {Priest}, \& {Mandrini}}]{Demoulin1996}
{D{\'e}moulin}, P., {Henoux}, J.~C., {Priest}, E.~R., \& {Mandrini}, C.~H.
  1996{\natexlab{a}}, \aap, 308, 643

\bibitem[{{D{\'e}moulin} {et~al.}(1996{\natexlab{b}}){D{\'e}moulin}, {Priest},
  \& {Lonie}}]{Demoulin1996a}
{D{\'e}moulin}, P., {Priest}, E.~R., \& {Lonie}, D.~P. 1996{\natexlab{b}},
  \jgr, 101, 7631, \dodoi{10.1029/95JA03558}

\bibitem[{{Downs} {et~al.}(2021){Downs}, {Warmuth}, {Long}, {Bloomfield},
  {Kwon}, {Veronig}, {Vourlidas}, \& {Vr{\v{s}}nak}}]{downs21}
{Downs}, C., {Warmuth}, A., {Long}, D.~M., {et~al.} 2021, \apj, 911, 118,
  \dodoi{10.3847/1538-4357/abea78}

\bibitem[{{Fan}(2017)}]{Fan2017}
{Fan}, Y. 2017, \apj, 844, 26, \dodoi{10.3847/1538-4357/aa7a56}

\bibitem[{{Finn} \& {Antonsen}(1985)}]{Finn1985}
{Finn}, J.~M., \& {Antonsen}, Thomas~M., J. 1985, Comments on Plasma Physics
  and Controlled Fusion, 9, 111

\bibitem[{{Forbes} {et~al.}(2018){Forbes}, {Seaton}, \& {Reeves}}]{Forbes2018}
{Forbes}, T.~G., {Seaton}, D.~B., \& {Reeves}, K.~K. 2018, \apj, 858, 70,
  \dodoi{10.3847/1538-4357/aabad4}

\bibitem[{{Hassanin} \& {Kliem}(2016)}]{Hassanin2016}
{Hassanin}, A., \& {Kliem}, B. 2016, \apj, 832, 106,
  \dodoi{10.3847/0004-637X/832/2/106}

\bibitem[{{Hassanin} {et~al.}(2022){Hassanin}, {Kliem}, {Seehafer}, \&
  {T{\"o}r{\"o}k}}]{Hassanin2022}
{Hassanin}, A., {Kliem}, B., {Seehafer}, N., \& {T{\"o}r{\"o}k}, T. 2022,
  \apjl, 929, L23, \dodoi{10.3847/2041-8213/ac64a9}

\bibitem[{{Isenberg} \& {Forbes}(2007)}]{Isenberg2007}
{Isenberg}, P.~A., \& {Forbes}, T.~G. 2007, \apj, 670, 1453,
  \dodoi{10.1086/522025}

\bibitem[{{Jin} {et~al.}(2017){Jin}, {Manchester}, {van der Holst}, {Sokolov},
  {T{\'o}th}, {Vourlidas}, {de Koning}, \& {Gombosi}}]{Jin2017}
{Jin}, M., {Manchester}, W.~B., {van der Holst}, B., {et~al.} 2017, \apj, 834,
  172, \dodoi{10.3847/1538-4357/834/2/172}

\bibitem[{{Jing} {et~al.}(2018){Jing}, {Liu}, {Lee}, {Ji}, {Liu}, {Xu}, \&
  {Wang}}]{Jing2018}
{Jing}, J., {Liu}, C., {Lee}, J., {et~al.} 2018, \apj, 864, 138,
  \dodoi{10.3847/1538-4357/aad6e4}

\bibitem[{{Kliem} \& {T{\"o}r{\"o}k}(2006)}]{kliem06}
{Kliem}, B., \& {T{\"o}r{\"o}k}, T. 2006, prl, 96, 255002,
  \dodoi{10.1103/PhysRevLett.96.255002}

\bibitem[{Kusano {et~al.}(2020)Kusano, Iju, Bamba, \& Inoue}]{Kusano2020}
Kusano, K., Iju, T., Bamba, Y., \& Inoue, S. 2020, Science, 369, 587,
  \dodoi{10.1126/science.aaz2511}

\bibitem[{{Linker} {et~al.}(2001){Linker}, {Lionello}, {Miki{\'c}}, \&
  {Amari}}]{Linker2001}
{Linker}, J.~A., {Lionello}, R., {Miki{\'c}}, Z., \& {Amari}, T. 2001, \jgr,
  106, 25165, \dodoi{10.1029/2000JA004020}

\bibitem[{{Linker} {et~al.}(2003){Linker}, {Miki{\'c}}, {Lionello}, {Riley},
  {Amari}, \& {Odstrcil}}]{linkeretal2003}
{Linker}, J.~A., {Miki{\'c}}, Z., {Lionello}, R., {et~al.} 2003, Physics of
  Plasmas, 10, 1971, \dodoi{10.1063/1.1563668}

\bibitem[{Liu \& Su(2021)}]{LiuT2021}
Liu, T., \& Su, Y. 2021, \apj, 915, 55, \dodoi{10.3847/1538-4357/ac013a}

\bibitem[{{Lugaz} {et~al.}(2011){Lugaz}, {Downs}, {Shibata}, {Roussev}, {Asai},
  \& {Gombosi}}]{lugaz11}
{Lugaz}, N., {Downs}, C., {Shibata}, K., {et~al.} 2011, \apj, 738, 127,
  \dodoi{10.1088/0004-637X/738/2/127}

\bibitem[{{Manchester} {et~al.}(2008){Manchester}, {Vourlidas}, {T{\'o}th},
  {Lugaz}, {Roussev}, {Sokolov}, {Gombosi}, {De Zeeuw}, \&
  {Opher}}]{manchester08}
{Manchester}, IV, W.~B., {Vourlidas}, A., {T{\'o}th}, G., {et~al.} 2008, \apj,
  684, 1448, \dodoi{10.1086/590231}

\bibitem[{{Miki{\'c}} {et~al.}(2013){Miki{\'c}}, {T{\"o}r{\"o}k}, {Titov},
  {Linker}, {Lionello}, {Downs}, \& {Riley}}]{Mikic2013a}
{Miki{\'c}}, Z., {T{\"o}r{\"o}k}, T., {Titov}, V., {et~al.} 2013, in American
  Institute of Physics Conference Series, Vol. 1539, American Institute of
  Physics Conference Series, ed. R.~B. J. C. S. C. H. E. J. G. W. G. G. L. E.
  M. E. M. N. P. J.~S. Gary P.~Zank, Joe~Borovsky \& O.~Verkhoglyadova (AIP
  Publishing), 42--45, \dodoi{10.1063/1.4810985}

\bibitem[{{Moore} {et~al.}(2001){Moore}, {Sterling}, {Hudson}, \&
  {Lemen}}]{Moore2001}
{Moore}, R.~L., {Sterling}, A.~C., {Hudson}, H.~S., \& {Lemen}, J.~R. 2001,
  \apj, 552, 833, \dodoi{10.1086/320559}

\bibitem[{{Patsourakos} \& {Vourlidas}(2009)}]{patsourakos09}
{Patsourakos}, S., \& {Vourlidas}, A. 2009, \apjl, 700, L182,
  \dodoi{10.1088/0004-637X/700/2/L182}

\bibitem[{{Priest}(1982)}]{Priest1982}
{Priest}, E.~R. 1982, {Solar magneto-hydrodynamics} (Dordrecht, Holland;
  Boston: D.~Reidel Pub.~Co.; Hingham), 469 p.

\bibitem[{{Priest} \& {D{\'e}moulin}(1995)}]{Priest1995}
{Priest}, E.~R., \& {D{\'e}moulin}, P. 1995, \jgr, 100, 23443,
  \dodoi{10.1029/95JA02740}

\bibitem[{{Savcheva} {et~al.}(2015){Savcheva}, {Pariat}, {McKillop},
  {McCauley}, {Hanson}, {Su}, {Werner}, \& {DeLuca}}]{Savcheva2015}
{Savcheva}, A., {Pariat}, E., {McKillop}, S., {et~al.} 2015, \apj, 810, 96,
  \dodoi{10.1088/0004-637X/810/2/96}

\bibitem[{{Savcheva} \& {van Ballegooijen}(2009)}]{Savcheva2009a}
{Savcheva}, A., \& {van Ballegooijen}, A. 2009, \apj, 703, 1766,
  \dodoi{10.1088/0004-637X/703/2/1766}

\bibitem[{{Somov} \& {Titov}(1985{\natexlab{a}})}]{Somov1985a}
{Somov}, B.~V., \& {Titov}, V.~S. 1985{\natexlab{a}}, \solphys, 95, 141,
  \dodoi{10.1007/BF00162642}

\bibitem[{{Somov} \& {Titov}(1985{\natexlab{b}})}]{Somov1985b}
---. 1985{\natexlab{b}}, \solphys, 102, 79, \dodoi{10.1007/BF00154039}

\bibitem[{{Syrovatskii}(1982)}]{Syrovatskii1982}
{Syrovatskii}, S.~I. 1982, \solphys, 76, 3, \dodoi{10.1007/BF00214126}

\bibitem[{{Titov}(2007)}]{Titov2007a}
{Titov}, V.~S. 2007, \apj, 660, 863, \dodoi{10.1086/512671}

\bibitem[{{Titov} {et~al.}(2018){Titov}, {Downs}, {Miki{\'c}}, {T{\"o}r{\"o}k},
  {Linker}, \& {Caplan}}]{Titov2018}
{Titov}, V.~S., {Downs}, C., {Miki{\'c}}, Z., {et~al.} 2018, \apjl, 852, L21.
\newblock \url{https://iopscience.iop.org/article/10.3847/2041-8213/aaa3da}

\bibitem[{Titov {et~al.}(2021)Titov, Downs, T{\"o}r{\"o}k, Linker, Caplan, \&
  Lionello}]{Titov2021}
Titov, V.~S., Downs, C., T{\"o}r{\"o}k, T., {et~al.} 2021, \apjs, 255, 9,
  \dodoi{10.3847/1538-4365/abfe0f}

\bibitem[{{Titov} {et~al.}(2002){Titov}, {Hornig}, \&
  {D{\'e}moulin}}]{Titov2002}
{Titov}, V.~S., {Hornig}, G., \& {D{\'e}moulin}, P. 2002, \jgr, 107, 1164,
  \dodoi{10.1029/2001JA000278}

\bibitem[{{T{\"o}r{\"o}k} {et~al.}(2004){T{\"o}r{\"o}k}, {Kliem}, \&
  {Titov}}]{torok04}
{T{\"o}r{\"o}k}, T., {Kliem}, B., \& {Titov}, V.~S. 2004, \aap, 413, L27,
  \dodoi{10.1051/0004-6361:20031691}

\bibitem[{T{\"o}r{\"o}k {et~al.}(2018)T{\"o}r{\"o}k, Downs, Linker, Lionello,
  Titov, Miki{\'c}, Riley, Caplan, \& Wijaya}]{Torok2018}
T{\"o}r{\"o}k, T., Downs, C., Linker, J.~A., {et~al.} 2018, \apj, 856, 75.
\newblock \url{http://stacks.iop.org/0004-637X/856/i=1/a=75}

\bibitem[{{Tziotziou, K.} {et~al.}(2014){Tziotziou, K.}, {Moraitis, K.},
  {Georgoulis, M. K.}, \& {Archontis, V.}}]{Tziotziou2014}
{Tziotziou, K.}, {Moraitis, K.}, {Georgoulis, M. K.}, \& {Archontis, V.} 2014,
  A\&A, 570, L1, \dodoi{10.1051/0004-6361/201424864}

\bibitem[{{van Ballegooijen}(2004)}]{vanBall2004}
{van Ballegooijen}, A.~A. 2004, \apj, 612, 519, \dodoi{10.1086/422512}

\bibitem[{{van Ballegooijen} \& {Martens}(1989)}]{vanBall1989}
{van Ballegooijen}, A.~A., \& {Martens}, P.~C.~H. 1989, \apj, 343, 971,
  \dodoi{10.1086/167766}

\bibitem[{{Zuccarello} {et~al.}(2012){Zuccarello}, {Meliani}, \&
  {Poedts}}]{Zuccarello2012}
{Zuccarello}, F.~P., {Meliani}, Z., \& {Poedts}, S. 2012, \apj, 758, 117,
  \dodoi{10.1088/0004-637X/758/2/117}

\end{thebibliography}



\end{document}